\documentclass[aps, prd, 10pt, twocolumn, eqsecnum, tightenlines, notitlepage, superscriptaddress, nofootinbib, preprintnumbers, floatfix]{revtex4-1}
\pdfoutput=1
\usepackage{epsfig,amsfonts,mathrsfs,amsmath,amssymb,graphicx,color,slashed}
\usepackage{amsmath,latexsym,amssymb,graphicx,slashed,hyperref,color,enumerate,url,etoolbox,cancel}
\usepackage[utf8]{inputenc}
\hypersetup{colorlinks,citecolor= nicegreen,linkcolor= nicered}
\definecolor{nicered}{rgb}{0.7,0.1,0.1}
\definecolor{nicegreen}{rgb}{0.1,0.5,0.1}

\def\Fermilab{Theoretical Physics Department, Fermilab, P.O. Box 500, Batavia, IL 60510, USA}
\def\Northwestern{Department of Physics and Astronomy, Northwestern University, Evanston, IL 60208, USA}
\def\Argonne{High Energy Physics Division, Argonne National Laboratory, Lemont, IL 60439, USA}
\def\Carleton{Ottawa-Carleton Institute for Physics, Carleton University, Ottawa, Ontario K1S 5B6, Canada}

\begin{document}

\title{Higgs Portal to Dark QED}

\author{Anirudh Krovi}
\affiliation{\Northwestern}
\author{Ian Low}
\affiliation{\Northwestern}
\affiliation{\Argonne}
\author{Yue Zhang}
\affiliation{\Carleton}
\affiliation{\Northwestern}
\affiliation{\Fermilab}

\date{\today}

\begin{abstract}
We introduce a light dark photon $A_\mu^\prime$ to the minimal Higgs portal model, by coupling the Higgs boson to "dark QED" containing fermionic dark matter, which gives rise to rich and interesting collider phenomenology. There are two prominent features in such a simple extension -- the Higgs boson could have decays into the long-lived dark photon through the  ``mono-$A^\prime$ channel,'' or into multiple collimated leptons via ``darkonium,'' depending on the mixing parameter of the $A_\mu^\prime$ with the visible photon. We  initiate a study on the possibility of probing the parameter space of the model in both the energy and the lifetime frontiers at the Large Hadron Collider.
\end{abstract}

\preprint{FERMILAB-PUB-19-472-T}

\maketitle 

\section{Introduction}

The discovery of the Higgs boson~\cite{Aad:2012tfa, Chatrchyan:2012xdj} provides a unique probe for many other puzzles in the Universe. One example is the nature of dark matter, which is among the most important open questions in modern science. If the dark matter is a particle, given that the Higgs boson couples to every fundamental particle (except, perhaps, the neutrino), it seems plausible that the Higgs boson could also interact with the dark matter.  The minimal model, where the only new ingredient is the dark matter particle, is  very simple and predictive, as the only free parameters are the dark matter mass and its coupling to the Higgs boson. Such  ``Higgs portal'' scenario  is quite appealing and has been widely explored \cite{Silveira:1985rk, Burgess:2000yq, Low:2011kp, Djouadi:2011aa, Cline:2013gha, Feng:2014vea, Beniwal:2015sdl, Arcadi:2019lka}. 

There have been a tremendous amount of experimental efforts to look for dark matter conclusively. These searches supply a handful of useful information on  properties of the dark matter, which in turn can be used to constrain the parameter space in the Higgs portal scenario. For example, direct detection of dark matter through the elastic scattering with nucleons provide robust bounds on the mass of the dark matter and its coupling to the Higgs boson. And if the dark matter is lighter than half the Higgs mass, the 125 GeV Higgs could decay into the dark matter invisibly, which can be constrained by measurements of the invisible decay width of the Higgs boson at the Large Hadron Collider (LHC). In this mass range, as it turned out, direct detection and invisible decay width constraints already exclude regions of parameter space giving rise to the observed relic density for a thermal dark matter particle~\cite{Low:2011kp}. In the viable region of parameter space, the interaction of the dark matter with the Higgs boson is so weak that the existence of a thermal dark matter would overclose the Universe. 

The overclosure constraint could be alleviated if addition annihilation channels of the dark matter exist, which necessitates going beyond the minimal Higgs portal scenario. One of the goals of the present work is to consider a simple extension of the minimal Higgs portal model, by including a light dark photon, which is a very popular hypothetical particle in ``dark sector'' scenarios~\cite{Pospelov:2007mp, Feng:2008ya, Essig:2009nc, An:2009vq, An:2015pva, Izaguirre:2017bqb, Tulin:2017ara, Krovi:2018fdr, Tsai:2018vjv} but less commonly considered in Higgs portal models. More specifically, we propose that the Higgs boson is the portal to dark QED, where the dark electron is a dark matter candidate. We will see that such a simple extension produces a rich set of collider phenomenology that could be explored at the Large Hadron Collider (LHC) at CERN. Indeed, the 125 GeV Higgs boson can serve as a useful probe not just for the dark matter, but also a more involved dark sector including a light dark photon. 

We point out two interesting features of a next-to-minimal dark side of the Higgs boson: 1) 
if the mixing of the dark photon with the visible photon is small, the dark photon will be long-lived and could be produced in the "mono-dark photon" channel via decays of the Higgs boson, 
and 2) 
if the dark photon has a sizeable mixing with the visible photon, the Higgs boson could decay promptly into bound states of dark matter, the darkonium, which subsequently decay back to multiple highly collimated charged leptons, the lepton-jets \cite{Baumgart:2009tn}.
In other words, the decay would be either boosted or long-lived. 

Exotic decays of the 125 GeV Higgs boson have been studied systematically using the ``simplified model'' approach~\cite{Curtin:2013fra} and there have been dedicated experimental efforts in this regard \cite{Alimena:2019zri}. It turns out that the specific decays into lepton jets we explore in this work has not been searched for experimentally \cite{Aad:2019tua}, although previous works on different mechanisms of the Higgs boson decaying into related final states can be found in Refs.~\cite{Falkowski:2010cm,Chang:2013lfa,Curtin:2014cca, Chang:2016lfq,Lu:2017uur,Izaguirre:2018atq}. On the other hand, the long-lived decay is relatively unexplored territory for the 125 GeV Higgs boson, except in Refs.~\cite{Alipour-Fard:2018lsf,Filimonova:2018qdc}.  We perform a preliminary study on the possible reach at the MATHUSLA detector~\cite{Curtin:2018mvb}, which is designed to measure long-lived particles produced at the LHC. 

This work is organized as follows. In Section \ref{sect:2} we introduce the model and present the Lagrangian of the Higgs portal to dark QED, followed by a discussion on the impact of the dark matter direct detection on the lifetime of the dark photon in Section \ref{sec:DD}. Then in Section \ref{sec:LHC} we consider the decay of the Higgs boson into the long-lived dark photon through the mono-$A^\prime$ channel at the MATHUSLA detector. Prompt decays into the lepton-jet are motivated and  
studied in Section \ref{sec:DMBS}. We  conclude in Section \ref{sect:con}.

\section{The Model}
\label{sect:2}

As a starting point, we consider the following Lagrangian for the Higgs portal to dark QED,
\begin{eqnarray}\label{eq:L0}
\begin{split}
\mathcal{L} = &\bar \chi i \cancel{D} \chi  - m_\chi \bar\chi \chi  + \frac{1}{\Lambda} \chi\chi^c \left(H^\dagger H - v^2/2 \right)  + {\rm h.c.} \\
& - \frac{1}{4} F'_{\mu\nu} F'^{\mu\nu} + \frac{1}{2} m_{A'}^2 A'_\mu A'^\mu - \frac{\kappa}{2} F'_{\mu\nu} F^{\mu\nu} \ .
\end{split}
\end{eqnarray}
Here $\chi$ is a Dirac fermion dark matter candidate and couples to the Standard Model (SM) only through the dimension-five interaction with the Higgs boson, which is suppressed by the cutoff scale $\Lambda$. In addition, $v=246\,$GeV is the Higgs vacuum expectation value.
We assume $\chi$ to be charged under a dark $U(1)_D$ gauge symmetry whose gauge boson is $V_\mu$. The dark $U(1)_D$ gauge boson could have kinetic mixing with the SM hypercharge gauge boson $B^\mu$, which at low energies manifests as a kinetic mixing between dark photon $A^\prime_\mu$ and the visible photon. In Eq.~(\ref{eq:L0}) $D_\mu = \partial_\mu + i g_D A'_\mu$. 
It is worth mentioning that the dark gauge symmetry $U(1)_D$ strictly forbids operators such as $\bar \chi LH$ which could otherwise destabilize our dark matter candidate $\chi$.

We define the ``dark Yukawa coupling'' between the Higgs boson and fermionic dark matter $\chi$ as
\begin{equation}
y_X \equiv \frac{v}{\Lambda} \ ,
\end{equation}
which, after electroweak symmetry breaking, is on the same footing as the Higgs-fermion couplings in the SM.

In this model, there are two ways for the dark sector to interact with the SM: via the Higgs portal or via the dark photon portal, which arises because of the kinetic mixing. We shall assume this kinematic mixing parameter $\kappa$ to be small enough so that the Higgs portal interaction offers the dominant dark matter production channel at the LHC, and $\kappa$ is only relevant for the dark photon to decay back to visible particles once produced on-shell.  However, even under the above assumption, a very light dark photon could still play an important role in the direct detection of dark matter. This constraint has a significant  impact on the allowed size of the kinematic mixing parameter $\kappa$ and the corresponding dark sector signals for LHC searches. We will elaborate on this important point in the upcoming section~\ref{sec:DD}. 

Before closing this section, we address the dark matter relic density  in this model.
Given the setup, there are two ways for the dark matter and its antiparticle to annihilate. One is through the Higgs portal, and the other through  annihilations into the dark photons, which cross section is not suppressed by the small kinetic mixing. 
For the sake of optimizing LHC signals, we will be interested in sizable (order one) dark sector gauge coupling values so that the annihilation cross section into dark photons is well above the values required for the thermal relic density. Therefore to arrive at the observed relic density one has to employ the asymmetric dark matter scenario \cite{Nussinov:1985xr, Kaplan:2009ag}, where the dark matter relic density is set by a primordial number density asymmetry between the $\chi$ and $\bar\chi$. 
As a byproduct of this assumption, all indirect detection constraints can be suppressed.

\section{Impact of Direct Detection on Dark Photon Decay Length}\label{sec:DD}

With a non-zero kinetic mixing $\kappa$, the dark matter scattering on nuclear targets could occur via $t$-channel dark photon exchange. At the nucleon level, the spin-independent scattering cross section is
\begin{equation}
\sigma_{SI}^{A'} = \frac{4\alpha_D \kappa^2 e^2 \mu_{N\chi}^{2}}{m_{A'}^4} \ ,
\end{equation}
where $\alpha_D=g_D^2/(4\pi)$ and $\mu_{N\chi}=m_\chi m_N/(m_\chi + m_N)$ is the reduced mass between dark matter and the target nucleon $N$ (can only be proton here).
The above expression is valid when dark photon mass is well above the typical momentum transfer.

With a very light (sub-GeV) dark photon, such a scattering cross section can be huge unless the kinetic mixing parameter $\kappa$ is very small. 
For dark matter mass above 10\,GeV or so, satisfying the current direct detection limits from XENON~\cite{Aprile:2018dbl}, LUX~\cite{Akerib:2016vxi} and PandaX-II~\cite{Cui:2017nnn} requires
\begin{equation}\label{eq:smallkappa}
\kappa \lesssim 10^{-9} \left( \frac{m_{A'}}{1\,\rm GeV} \right)^2 \left( \frac{0.5}{\alpha_D} \right)^{1/2} \ ,
\end{equation}
which applies to the  model introduced in Eq.~(\ref{eq:L0}). With such a small $\kappa$, the dark photon will be long lived if it has to decay back to the visible sector. Its decay length with an order one boost factor is
\begin{equation}\label{eq:decaylength}
c\tau_{A'} \simeq 10^{7}\, {\rm cm}\, \left( \frac{10^{-9}}{\kappa} \right)^2 \left( \frac{1\,{\rm GeV}}{m_{A'}} \right) \ .
\end{equation}
Therefore, if the dark photon is produced at colliders, it will appear as a long-lived particle. This argument holds for dark photon lighter than $\sim20\,$GeV and regardless how the dark photon is produced. In subsection~\ref{sec:LLDP}, we will explore the possibility of producing the dark photon from the mono-dark photon channel in the Higgs decays and how to search for it using the recently proposed MATHUSLA detector for LHC run 3 and beyond~\cite{Curtin:2018mvb}.

Alternatively, the dark sector structure could be more involved as in the real world. In section~\ref{sec:DMBS}, we will consider a scenario where dark-matter-nucleon scattering via the dark photon becomes inelastic \cite{TuckerSmith:2001hy}. With a large enough mass splitting between the dark matter and its heavier partner, the severe constraint Eq.~(\ref{eq:smallkappa}) can be evaded. As a result, the dark photon could decay promptly at the LHC. This will be studied in more detail in  Section \ref{sec:DMBS}.

In both the long-lived and the prompt decays scenarios, even if the dark photon mediated scattering is suppressed, there's still the contribution from the Higgs exchange, which gives rise to the spin-independent  cross section at nucleon level
\begin{equation}\label{eq:higgsex}
\sigma_{SI}^{h} = y_X^2 \frac{4 f^2 m_N^2 \mu_{N\chi}^{2}}{\pi v^2 m_h^4} \ ,
\end{equation}
where $N$ can be proton or neutron in this case. In the following we will use Eq.~(\ref{eq:higgsex}) to place constraints on the dark Yukawa coupling $y_X$ using current limits from the direct detection experiments~\cite{Aprile:2018dbl, Akerib:2016vxi, Cui:2017nnn}, which correspond to the blue shaded exclusion regions in Figs.~\ref{fig:Long_Lived} and \ref{fig:Multi_Lepton}.

\section{Higgs decay to dark sector: long-lived dark photon case}\label{sec:LHC}

\subsection{Invisible decay of the Higgs boson}\label{sec:invisible}

For $m_\chi<m_h/2$, the Higgs boson in the model can decay into $\chi, \bar\chi$, which contributes to the invisible width of the Higgs boson decay.
The corresponding partial decay rate is
\begin{equation}
\Gamma_{h\to\chi\bar\chi} = y_X^2 \frac{\left(m_{h}^{2}-4m_{\chi}^{2}\right)^{3/2}}{8\pi m_{h}^{2}} \ .
\end{equation}
Such an invisible decay branching ratio is constrained indirectly from the global fit to the Higgs production and decay rates into the SM final states at the LHC. The present bound is~\cite{Tanabashi:2018oca}
\begin{equation}
{\rm Br}_{h\to {invisible}} < 24\% \ ,
\end{equation}
which translates to the purple shaded exclusion regions in Figs.~\ref{fig:Long_Lived} and \ref{fig:Multi_Lepton}.

\subsection{Higgs decay into dark matter and long-lived dark photon}\label{sec:LLDP}

\begin{figure}[t]
\centerline{\includegraphics[width=0.45\textwidth]{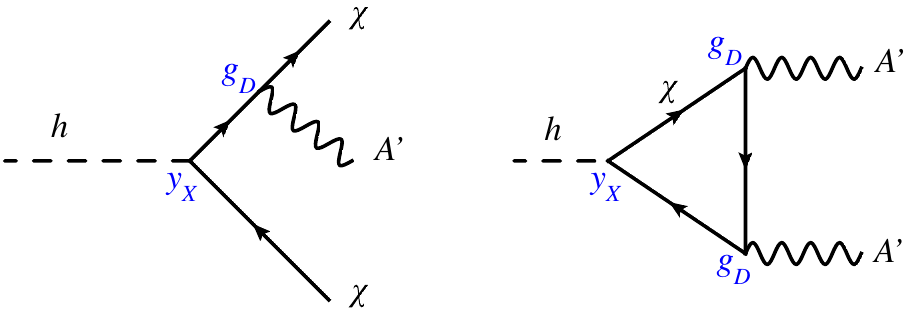}}
\caption{Feynman diagrams for dark photon production via the Higgs portal interaction.}\label{fig:Fey0}
\end{figure}

The next Higgs decay channel we study contains a dark photon in the final state. This could be interesting, as argued in section~\ref{sec:DD}, when the dark photon is long lived in the minimal model in order to satisfy the dark matter direct detection bounds. Because the kinetic mixing parameter is constrained to be very tiny (see Eq.~\ref{eq:smallkappa}), the direct production of dark photon from quark-anti-quark annihilation is not efficient. 
Instead, dark photon is dominantly produced through decays of the Higgs boson.

Here, we take into account of two Higgs decay channels which produce the dark photon. 
The first channel is $h\to \chi\bar\chi A'$, as shown in Fig.~\ref{fig:Fey0} (left). 
Here the dark matter particles in the final state appear as missing energy at colliders, 
so it is a ``mono-dark-photon'' process.
The corresponding decay rate is proportional to $y_X^2 \alpha_D$, but is a rather complicated function of $m_\chi$ and $m_{A'}$. Therefore, we do not show its full analytic expression here. In particular, in the limit of massless $\chi$ and $A'$, this decay rate features the infrared and collinear divergences, which must be regularized by adding it to the $h\to \chi\bar\chi$ decay rate calculated up to next-to-leading order in $\alpha_D$.
In practice, we evaluate the mono-dark photon decay rate numerically using {\tt MadGraph}~\cite{Alwall:2011uj}.

\begin{figure}[t]
\centerline{\includegraphics[width=0.46\textwidth]{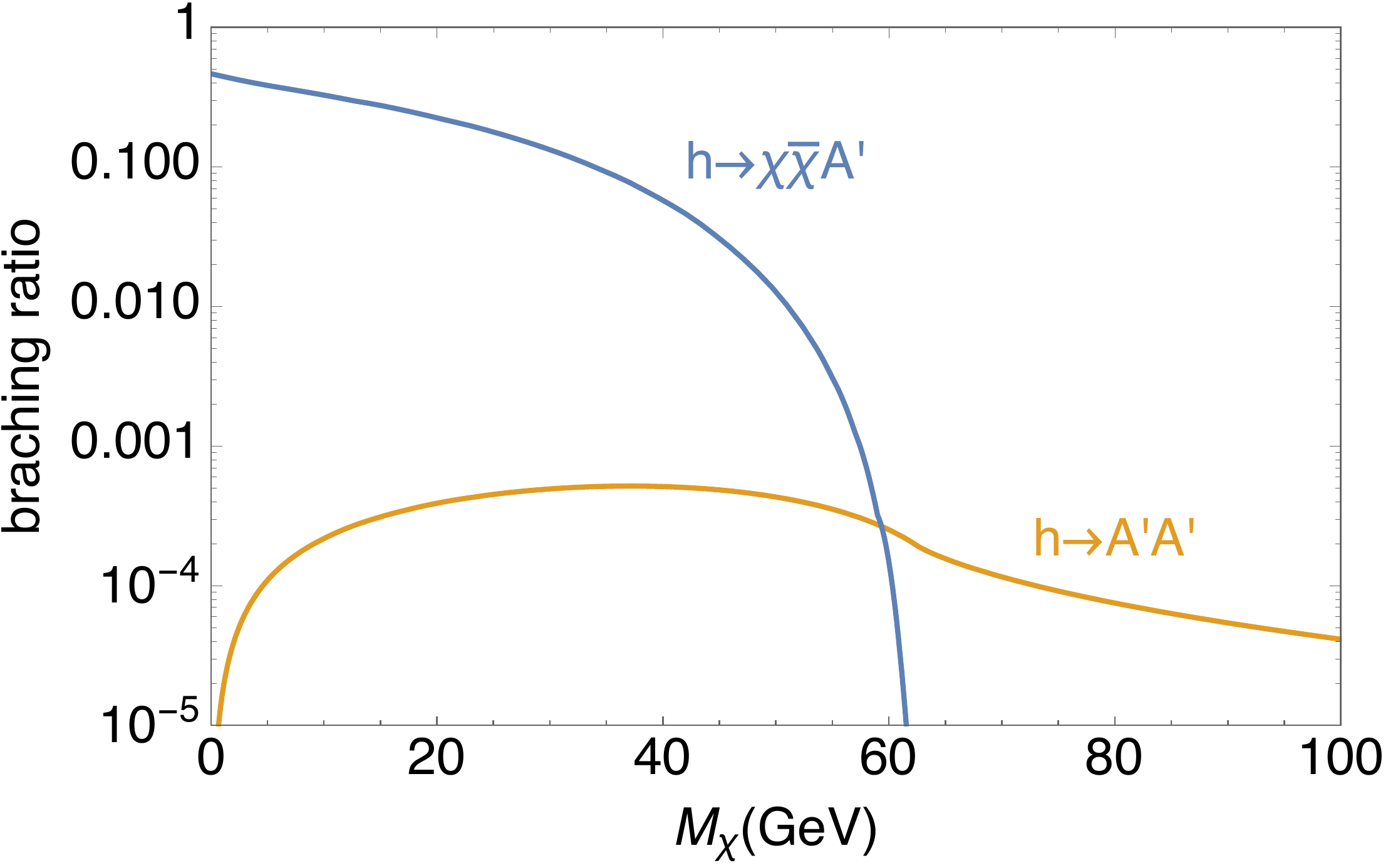}}
\caption{Higgs decay branching ratios as a function of dark matter mass, with other model parameters fixed, $\alpha_D=0.5$, $m_{A'}=0.5\,$GeV, $y_X=0.01$.}\label{fig:decaycompete}
\end{figure}

The second Higgs decay channel we consider is $h\to A'A'$ which is generated at loop level, as shown in Fig.~\ref{fig:Fey0} (right).
The decay rate for this channel is \cite{Djouadi:2005gi}
\begin{eqnarray}
&&\Gamma_{h\to A'A'} = y_X^2 \frac{\alpha_D^2 m_\chi^2}{16\pi^3 m_h} \left| A\left( \frac{m_h^2}{4m_\chi^2} \right) \right|^2 \ , \\
&&A(x) = x + (x-1) f(x), \nonumber\\
&&f(x) = \left\{ \begin{array}{lr}
\arcsin^2(\sqrt{x}), & x\leq1 \\
-\frac{1}{4}\left( \log\frac{1+\sqrt{1-x^{-1}}}{1-\sqrt{1-x^{-1}}} - i\pi \right)^2, & x>1
\end{array} \right.\nonumber
\end{eqnarray}
In practice, we find that the decay rate for $h\to \chi\bar\chi A'$ is much larger than that for $h\to A'A'$ when $m_\chi < m_h/2$, whereas the latter dominates when $m_\chi \gtrsim m_h/2$ because the former become kinematically forbidden.
In Fig.~\ref{fig:decaycompete}, we plot the two decay rates and show their interplay for a set of the model parameters.
In practice, we find that given the LHC projected luminosity, only the $h\to \chi\bar\chi A'$ decay will give a competitive limit, as will be show below.

\begin{figure}[t]
\centerline{\includegraphics[width=0.46\textwidth]{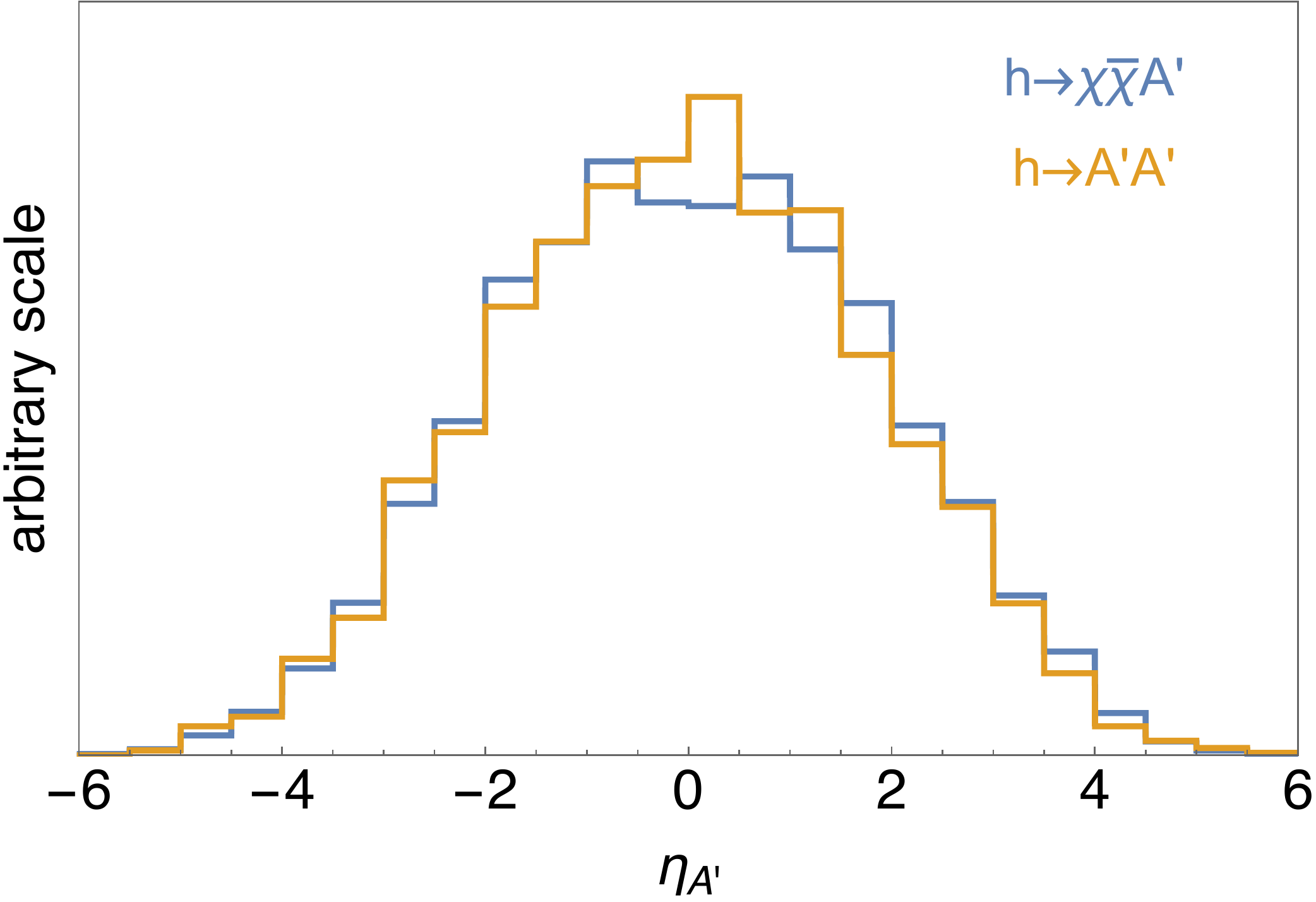}}
\caption{Pseudo-rapidity distribution for the $h\to \chi\bar\chi A'$ and $h\to A'A'$ processes. Note the relative height between the two distributions does not reflect the actual rates.}\label{fig:Eta_Dist}
\end{figure}

We simulate mono-$A_\mu^\prime$ process for 13 TeV LHC using {\tt MadGraph}~\cite{Alwall:2011uj}. The pseudo-rapidity distributions of the dark photon in the laboratory frame are shown in Fig.~\ref{fig:Eta_Dist} for fixed dark matter and dark photon masses, $m_\chi=10\,$GeV and $m_{A'}=0.5\,$GeV, but we find the features are quite generic throughout the model parameter space.
The majority of the produced dark photons travel in the central directions instead of being forward with respect to the beam. In addition, as estimated in Eq.~(\ref{eq:decaylength}), the typical dark photon decay length is much longer than any of the existing and proposed detectors associated with the LHC experiment. These considerations lead us to detectors located in the non-forward directions and having as large a volume as possible. Among the proposals currently being considered, we find that MATHUSLA~\cite{Curtin:2018mvb} works the best in fulfilling these requirements. We also notice that for detecting the decay product of the dark photon, the minimal transverse momentum in the event selection at MATHUSLA is quite low (around GeV or less)~\cite{Curtin:2018mvb}.
Most of the dark photons from Higgs boson decay in our models are energetic enough  to pass this cut.

In order to calculate the number of signal events detected by MATHUSLA, we define three efficiency factors:
\begin{itemize}
\item Among all the simulated events, we first impose an angular distribution cut selecting the produced dark photons that point within the solid angle spanned by the MATHUSLA detector. The fraction of events passing this cut is defined as the efficiency factor $\epsilon_1$. 
\item Next, the dark photon must decay within the MATHUSLA detector. Assuming that there are $N_1$ events that pass the angular cut. For each event, the dark photon has a boost factor $\gamma = E/m_{A'}$, and its trajectory intersects with the detector at distance $x_{\rm in}$ and $x_{\rm out}$ from the primary interaction point. 
The corresponding efficiency factor $\epsilon_2$ can then be calculated as
\begin{eqnarray}
\epsilon_2 &=& \frac{1}{N_1} \sum_{a=1}^{N_1} \left[ \exp\left(- \frac{x_{\rm in}^a}{\gamma_a c \tau_{A'}}\right) \right.\nonumber\\
       &&\qquad\qquad \left. - \exp\left(- \frac{x_{\rm out}^a}{\gamma_a c \tau_{A'}}\right) \right] \ .
\end{eqnarray}
In the limit $\gamma c\tau\gg x_{\rm out}- x_{\rm in}$, the product $\epsilon_1 \epsilon_2$ is proportional to the volume of detector.
\item Finally, the efficiency factor $\epsilon_3$ accounts for the fraction of events pass the three momentum cut on the decay products of the dark photon. 
In our analysis, we take into account of $e^\pm$ and $\mu^\pm$ in the final states, and the corresponding cuts are
$|\vec{p}|_{e^\pm}>1\,$GeV, $|\vec{p}|_{\mu^\pm}>0.2\,$GeV. 
\end{itemize}
With the dark photon mass benchmark used above ($m_{A'}=0.5\,$GeV), its decay branching ratios into $e^+e^-$ and $\mu^+\mu^-$ pairs are both $\sim 40\%$.
These branching ratios decrease for higher dark photon masses, when other hadronic decay channels open up kinematically.
In presenting our results below, we will scan over the dark photon mass.
A cut flow table is shown in Table~\ref{table:0} for a set of the model parameters.

\begin{table}[t]
\begin{center}
	\begin{tabular}{ |c|c|c| } 
		\hline
		 Efficiency & $h\to \chi\bar\chi A'$ & $h\to A' A'$\\ 
		 \hline		 
		$\epsilon_1$ &0.037 & 0.0762 \\
		\hline 
		$\epsilon_2$ & 1.14$\times10^{-4}$ & $5.92\times10^{-6}$\\ 
		\hline
		$\epsilon_3$  & 0.66 & 0.77 \\
		\hline
		\end{tabular}
	    \end{center}
\caption{Efficiency factors for passing each selection cut in using MATHUSLA to hunt long-lived dark photon, with a set of dark sector parameters, $m_{\chi}$ = 20\,GeV and $m_{A'}$ = 0.5\,GeV. 
See the text for the detailed definitions of $\epsilon_{1,2,3}$.
}
\label{table:0}
\end{table}

\begin{figure}[t]
\centerline{\includegraphics[width=0.46\textwidth]{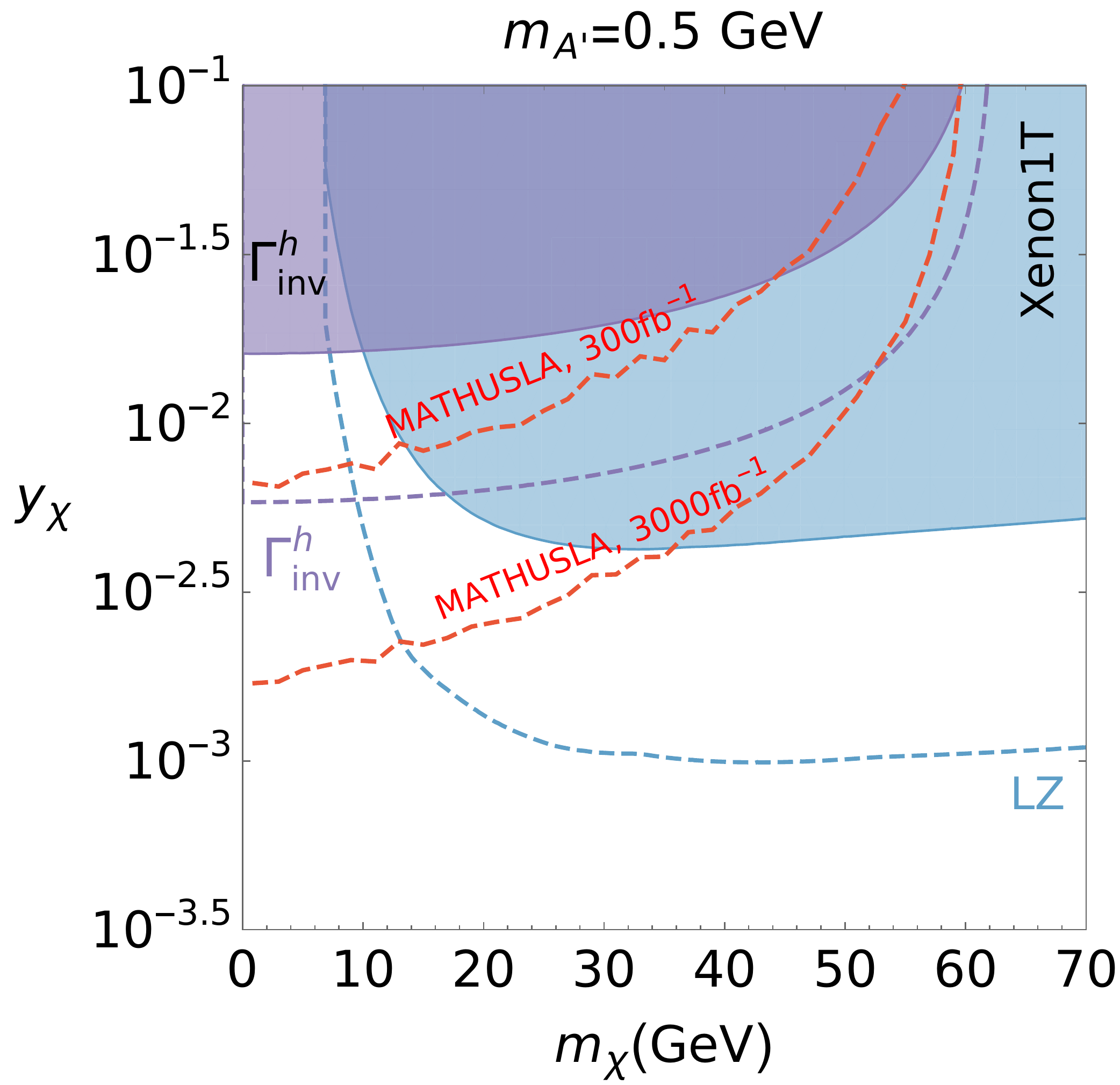}}
\centerline{\includegraphics[width=0.445\textwidth]{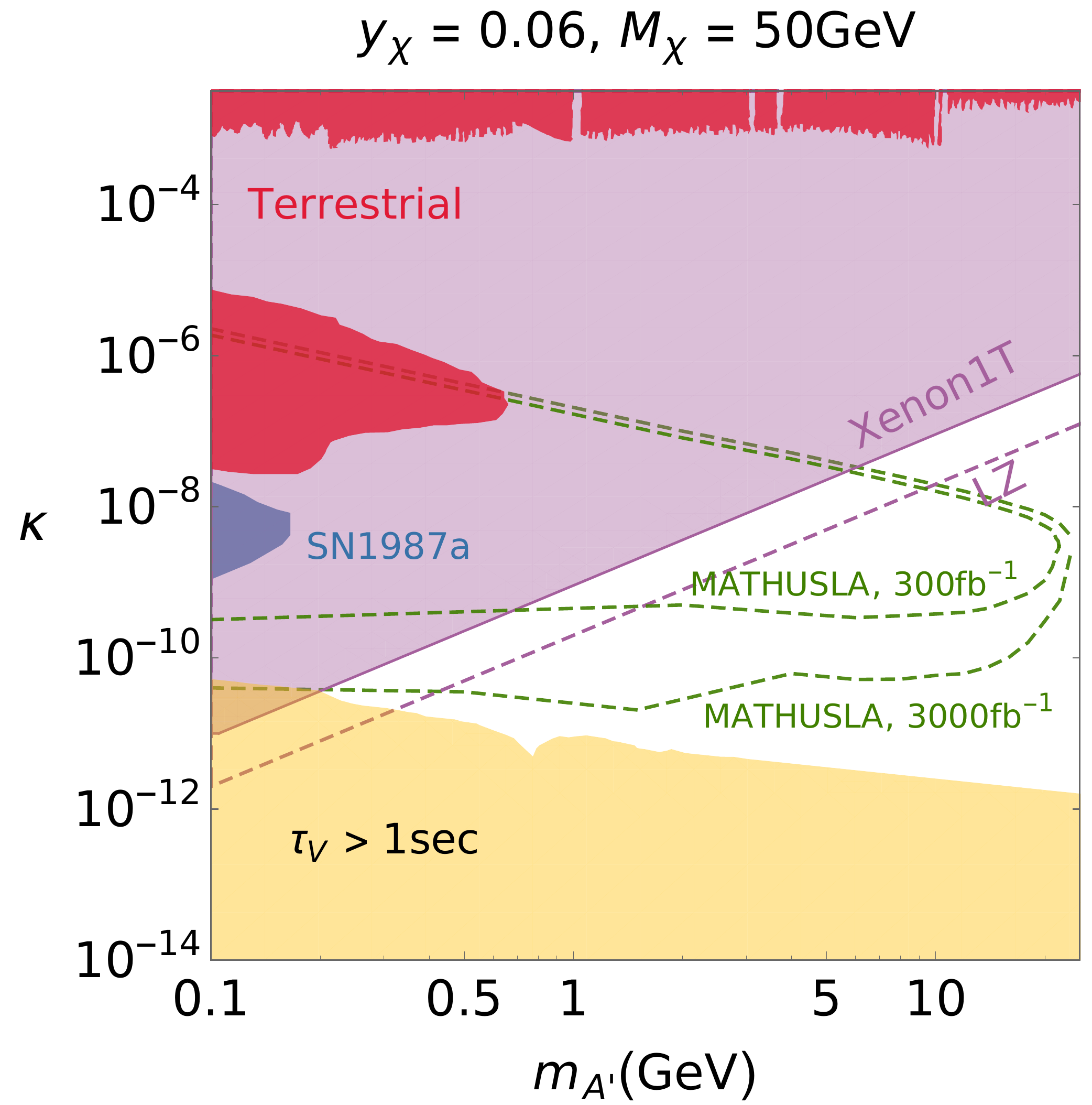}}
\caption{Long-lived dark photon search using MATHUSLA (expected reach with future LHC luminosities) versus the existing and projected constraints from dark matter direct detection and Higgs invisible decay constraints. In the upper plot, the reach is shown in the $y_X$ versus $m_\chi$ plane with other parameters fixed, $m_{A'} = 0.5\,$GeV, $\alpha_D=0.5$ and $\kappa=10^{-9}$; whereas in the lower plot, the reach is shown in the $\kappa$ versus $m_{A'}$ plane, with $m_{\chi}=50\,$GeV, $\alpha_D=0.5$ and $y_X=6\times10^{-2}$.}\label{fig:Long_Lived}
\end{figure}

The total number of signal events expected in the MATHUSLA detector is then given by
\begin{equation}
N_{\rm signal} = \mathcal{L} \sigma(gg\to h) {\rm Br} (h\to\chi\bar\chi) \epsilon_1 \epsilon_2 \epsilon_3 \ .
\end{equation}
We scan over the parameter space of the model and derive the region that could potentially be covered. The results are shown in Fig.~\ref{fig:Long_Lived}, where the red contours correspond to 3 signal events and zero background, which is the 95\% C.L. limit using Poisson statistics, for a given LHC integrated luminosity $\mathcal{L}$. 

In the upper plot, we hold $m_{A'} = 0.5\,$GeV, $\alpha_D=0.5$ and $\kappa=10^{-9}$ fixed and compare the MATHUSLA reach with the existing limits from Higgs invisible decay searches and dark matter direct detection. Interestingly, we find that MATHUSLA could probe regions with dark Yukawa coupling $y_X$ as small as $\sim 10^{-3}$. The limit gets weaker for heavier dark matter because of the phase space suppression in the partial Higgs decay rate. 

In the lower plot of Fig.~\ref{fig:Long_Lived}, we show the corresponding MATHUSLA reach in the $\kappa$ versus $m_{A'}$ parameter space, holding $m_{\chi}=50\,$GeV, $\alpha_D=0.5$ and $y_X=0.06$ fixed. This is to be compared with the existing visibly-decaying dark photon decay constraints. The regions in pink, labeled terrestrial, correspond to constraints coming from all terrestrial experiments looking for dark photons. This includes constraints from $e^+e^-$ colliders, electron and proton beam dump and experiments studying precision meson physics~\cite{Alexander:2016aln, Bauer:2018onh}. The region in blue corresponds to the constraints coming from the collapse of SN1987a, where the production of dark photons is constrained, given that it is known that most of the energy emitted is in the form of neutrinos~\cite{Chang:2016ntp}. In addition, since it is required that the dark photon decay in a short enough time such that it does not affect the standard Big Bang Nucleosynthesis picture, we use the often adopted condition that the dark photon decay within 1 second~\cite{Kaplinghat:2013yxa}. This constraint is plotted in yellow.
Interestingly, we find that the existence of a Higgs portal allows us to probe the region of parameter space where the dark photon is heavy (well above GeV scale) and long-lived.

\section{Higgs decay to dark sector: prompt dark photon case}\label{sec:DMBS}

In the previous section, we have studied the Higgs portal to dark QED and concentrated on a minimalistic setup, where null results from the direct detection of dark matter imply that the dark photon must have a very small mixing with the visible photon and therefore long lived at colliders. In this section, we add new ingredients to the minimal setup where the direct detection constraints are loosened. This results in new collider signatures at the LHC.

A well known example that could suppress the dark photon contribution to direct detection without a small $\kappa$ is the inelastic dark matter scenario~\cite{TuckerSmith:2001hy}. Namely, one could add a $U(1)_D$ breaking Majorana mass terms $(\Delta m/4) (\bar\chi \chi^c + \bar\chi^c \chi)$ to the Lagrangian in Eq.~(\ref{eq:L0}), where $\chi^c$ is the charge conjugation field of $\chi$.
This splits the Dirac fermion $\chi$ into two Majorana mass eigenstates, $\chi_{1,2}$, with mass difference equal to $\Delta m$. 
After introducing the Majorana mass, the theory still has an unbroken $\mathbb{Z}_2$ symmetry so that the lighter state $\chi_1$ can be stable and serve as the dark matter candidate.
Meanwhile, the fermion-dark-photon interaction term becomes off-diagonal, and the dark matter scattering process becomes endothermic.
If the mass splitting $\Delta m$ is larger than the maximal energy transfer, the direct detection constraints can be evaded.
As a result, we are allowed to consider larger values of $\kappa$ where the dark photon decay length is much shorter and its decay could be prompt.\footnote{In this case the upper bound on $\kappa$ come from dark photon search in low energy experiments. For $m_{A'}\lesssim 10\,$GeV, the bound is only $\kappa\lesssim 10^{-3}$~\cite{Alexander:2016aln}.\label{footnote1}} 

This possibility, however, is incompatible with our asymmetric dark matter assumption. Because Majorana mass break the global $U(1)_D$ symmetry,
nothing forbids two $\chi_1$ particles to annihilate into dark photons, which could then decay into SM particles.
It is further shown~\cite{Zhang:2016dck, Blennow:2016gde} that the annihilation and self interaction among the Majorana $\chi_1$ particles still feature 
the Sommerfeld enhancement effect when $\alpha_D > v,\, m_{A'}/m_\chi,\, \sqrt{\Delta m/m_\chi}$ ($v$ is the relative velocity between two dark matter particles). 
In this case, if $\chi_1$ accounts for all the observed dark matter relic density, the model suffers from severe indirect detection constraints~\cite{Bringmann:2016din}, especially for large $\alpha_D$ values where the collider search prospects are optimized.

Given the above considerations, we find a bit more model building is necessary in order to make the dark matter "inelastic" for the purpose of evading the direct detection constraints, while keeping 
it asymmetric for surviving  the strong indirect detection constraints.

\subsection{Model of asymmetric inelastic dark matter}

It is possible to achieve the above goal in a model  with two Dirac fermions in the dark sector, $\chi$ and $\psi$, which are oppositely charged under the $U(1)_D$. Their gauge interactions with the dark photon take the form,
\begin{eqnarray}
\mathcal{L}_{gauge} = g_D \left( \bar \chi \gamma^\mu \chi - \bar\psi \gamma^\mu \psi \right) A_\mu' \ .
\end{eqnarray}
We introduce the following mass terms
\begin{equation}
\mathcal{L}_{mass} = m_\chi (\bar \chi \chi + \bar \psi \psi) + \frac{\Delta m}{2} (\bar \chi \psi^c  + \bar\psi^c \chi) \ ,
\end{equation}
where the mass parameter $\Delta m$ breaks the $U(1)_D$ gauge symmetry thus is related to (part of) the dark photon mass.
The above mass terms could be diagonalized with the following field redefinition
\begin{eqnarray}
\chi \to \frac{\mu_D - i e_D}{\sqrt2}, \ \ \ \psi \to \frac{\mu_D + i e_D}{\sqrt2} \ ,
\end{eqnarray}
where $e_D, \mu_D$ are the two Dirac fermion mass eigenstates in the dark sector.
We name them in analogy to the electron and muon in our sector.
In the new basis, the above mass and gauge interaction terms become
\begin{equation}\label{eq:twoflavorgauge}
\begin{split}
\mathcal{L} = & \ (m_\chi - \Delta m/2) \bar e_D e_D + (m_\chi + \Delta m/2) \bar \mu_D \mu_D \\
& - i g_D \left(\bar \mu_D \gamma^\mu e_D - \bar e_D \gamma^\mu \mu_D \right) A_\mu' \ ,
\end{split}
\end{equation}
Clearly, the gauge interaction vertex involving $e_D, \mu_D$ and the dark photon is purely ``flavor'' off-diagonal -- different from the QED in the visible sector.
This resulting Lagrangian possesses an additional unbroken global $U(1)$ symmetry (the analogy of lepton number) where $e_D$ and $\mu_D$ (or $\chi$ and $\psi$) carry the same charge~\cite{TsaiY}.
This way, one can realize an asymmetric relic abundance for the dark matter candidate $e_D$ in this model, while still suppressing the dark photon contribution to direct detection with a large enough $\Delta m$.

\subsection{Turning on the Higgs portal}

In this model, we could turn on a Higgs portal interaction, for example,
\begin{equation}\label{eq:newHiggsportal}
\begin{split}
\mathcal{L}_{portal} &= \frac{1}{\Lambda} (H^\dagger H - v^2/2) \bar \chi \chi \\
&= \frac{v}{2\Lambda} h (\bar \mu_D + i \bar e_D) (\mu_D - i e_D) \ .
\end{split}
\end{equation}
This opens up four new decay modes of the Higgs boson, $h\to e_D^+ e_D^-,\ e_D^+ \mu_D^-,\ \mu_D^+ e_D^-,\ \mu_D^+ \mu_D^-$.
If the mass splitting between $\mu_D$ and $e_D$ is small enough so that $\mu_D$ is long-lived, 
all the four decay channels of the Higgs boson could appear as invisible final states at LHC detectors. 
In this case, the Higgs invisible decay constraint discussed in subsection~\ref{sec:invisible} can still set limit on the 
dark Yukawa coupling $y_X=v/\Lambda$, which is defined the same as before.

\subsection{Higgs decay into darkonium}

Next, we explore possible bound states from dark sector gauge interactions, the darkonium, and 
their roles in Higgs boson exotic decays at the LHC.

\begin{figure}[t]
\centerline{\includegraphics[width=0.2\textwidth]{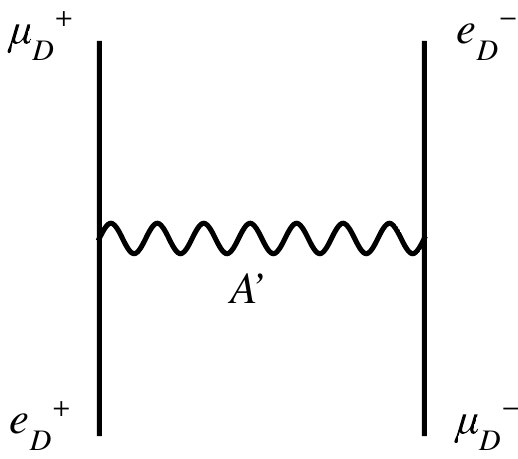}}
\caption{Feynman diagram contributing to the potential in Eq.~(\ref{eq:potential}).}\label{fig:Feypot}
\end{figure}

Interestingly, the gauge interaction Eq.~(\ref{eq:twoflavorgauge}) in the above two-flavor model 
allows two-body bound states to exist which is a linear combination of $\left|e_D^+ \mu_D^- \right\rangle$ and $\left|\mu_D^+ e_D^- \right\rangle$, two degenerate composite states. In this basis, the two-body interacting potential takes the following matrix form
\begin{equation}\label{eq:potential}
V(r) = \begin{pmatrix}
 0 & \frac{\alpha_D}{r}e^{- m_{A'} r} \\
 \frac{\alpha_D}{r} e^{- m_{A'} r} & 0 
\end{pmatrix} \ ,
\end{equation}
which arises from the dark photon exchange represented by the  Feynman diagram in Fig.~\ref{fig:Feypot}.
%
\begin{figure}[t]
\centerline{\includegraphics[width=0.45\textwidth]{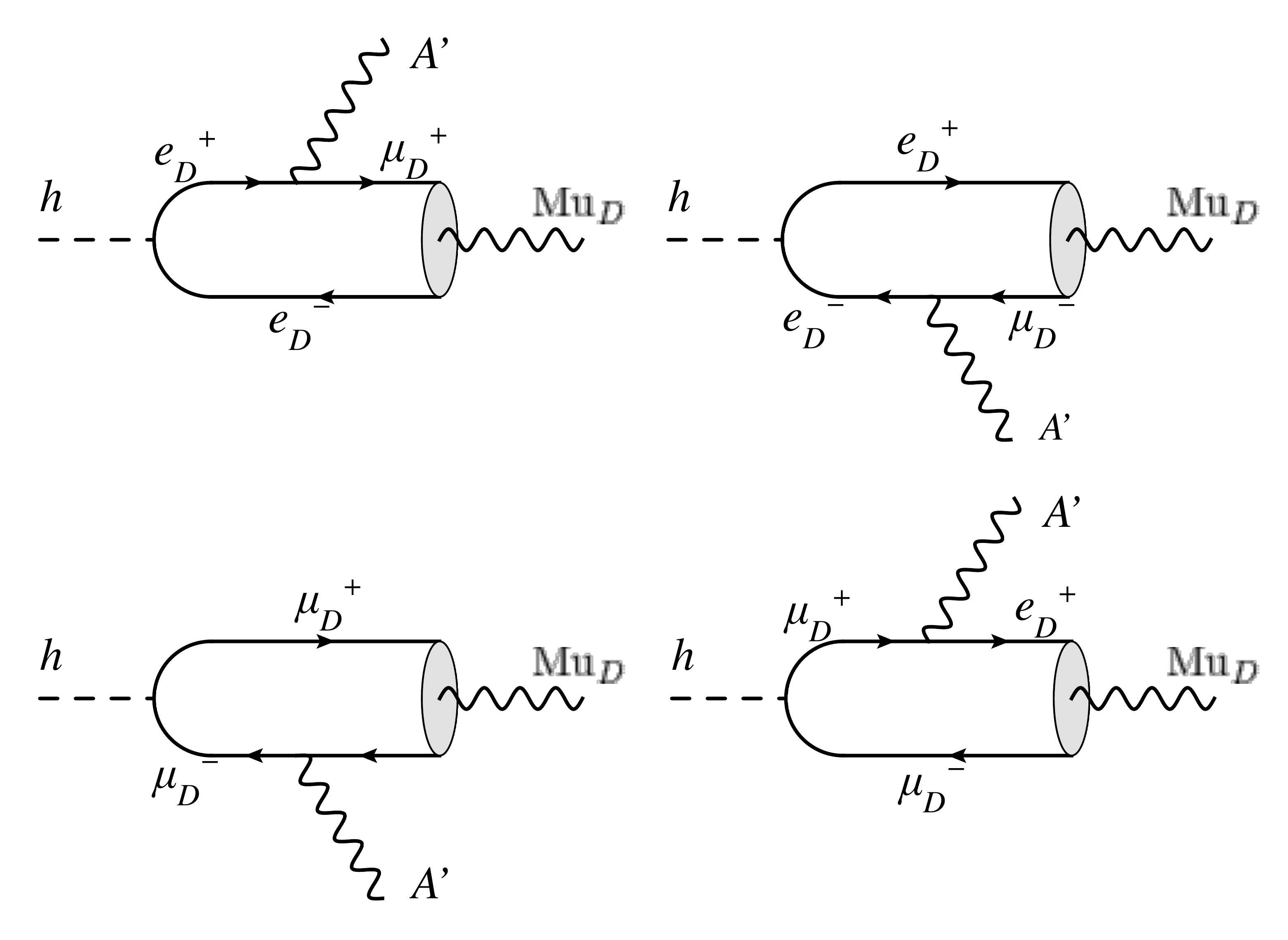}}
\caption{Feynman diagrams for Higgs boson decay into dark sector bound states, $h\to {\rm Mu}_D A'$.}\label{fig:Fey1}
\end{figure}
%
After diagonalizing this matrix, we find that
\begin{equation}\label{eq:state}
\frac{1}{\sqrt2}\left(\left|e_D^+ \mu_D^- \right\rangle + \left|\mu_D^+ e_D^- \right\rangle \right)
\end{equation}
has an attractive potential thus could form bound states, whereas the potential for the orthogonal state is repulsive.
This is the analogue of muonium in the visible sector.
In the remainder of this section, we will explore the the possible decay of the Higgs boson into such a dark muonium ground state.
We focus on the spin one ($J=S=1$, $L=0$) bound state ${\rm Mu}_D$ and a new Higgs boson decay channel $h\to {\rm Mu}_D A'$
through the Higgs portal interaction in Eq.~(\ref{eq:newHiggsportal}). The leading order Feynman diagrams are shown in Fig.~\ref{fig:Fey1}.
In contrast, the Higgs decay into a $J=S=L=0$ state plus one dark photon is forbidden due to $C$-parity conservation.\footnote{The dark muonium states have definite $C$-parities given by $(-1)^{L+S}$, which is even for the $S=0$ ground state. The parity of this state is odd thus it cannot mix with the Higgs boson unless one turns on another coupling $(H^\dagger H)\bar\chi i\gamma_5 \chi$ in addition to Eq.~(\ref{eq:newHiggsportal}). We do not consider this possibility in this work.}

In order for the dark muonium state to be actually formed, the lifetime of ${\rm Mu}_D$ should be longer than  the bound state lifetime, the time scale required for forming the bound state. This can be achieved by having a small enough mass splitting $\Delta m$ (but still large enough to suppress direct detection constraints, see~\cite{Krovi:2018fdr} for a quantitative discussion), which we assume to be the case throughout the discussions below.

We calculate partial decay rate for $h\to {\rm Mu}_D A'$, 
\begin{widetext}
\begin{equation}\label{eq:mudarate}
\Gamma_{h\to {\rm Mu}_D A'} = y_X^2
\frac{4 \alpha_D \vert\psi(0)\vert^{2} \left[ m_{A'}^4 - 2 m_{A'}^2 \left(m_h^2 - 8 m_\chi^2\right) + \left(m_h^2 - 4 m_\chi^2\right)^2 \right] \sqrt{ \left[ m_h^2 - \left(m_{{\rm Mu}_D} + m_{A'} \right)^2\right] \left[ m_h^2 - \left(m_{{\rm Mu}_D} - m_{A'} \right)^2\right]}}
{m_{h}^{3} m_{\chi} (m_{h}^{2}+m^{2}_{A'}-4m_{\chi}^{2})^{2}}  \ ,
\end{equation}
\end{widetext}
where we neglected the $e_D, \mu_D$ mass difference. The mass of the bound state, $m_{{\rm Mu}_D}$, is equal to $2m_\chi$ minus the binding energy~\cite{Krovi:2018fdr},
\begin{equation}
BE \simeq \frac{1}{4} \alpha_D m_\chi \left( 1 - \pi^2 m_{A'} a_0/12 \right)^2 ,
\end{equation}
where $a_0 = 2/(\alpha_D m_\chi)$ is the Bohr radius.
In Eq.~(\ref{eq:mudarate}) $\vert\psi(0)\vert$ is the bound state wave function at the origin. 
For a non-zero dark photon mass, the Yukawa potential problem could be evaluated numerically, or approximated with a Hulth\'en potential, which gives~\cite{Krovi:2018fdr}
\begin{equation}
\vert\psi(0)\vert \simeq \sqrt {[4- (\pi^2/6)^2 m_{A'}^2 a_0^2]/(4\pi a_0^3)} \ .
\end{equation}
In the Coulomb limit ($m_{A'}\to0$), we have $\vert\psi(0)\vert \propto (\alpha_D m_\chi)^{3/2}$, and the above decay rate becomes approximately 
\begin{equation}\label{eq:simplifiedBr}
\Gamma_{h\to{\rm Mu}_D A'} \simeq y_X^2 \frac{\alpha_D^4 m_\chi^2 \left(m_h^2 - m_{{\rm Mu}_D}^2 \right)}{2\pi m_h^3} \ .
\end{equation} 
Because this decay rate depends on the fourth power of $\alpha_D$, we find that $\alpha_D$ cannot be much smaller than one otherwise the Higgs branching ratio in this channel would be too small.

The ${\rm Mu}_D$ bound state, unlike the muonium in the visible sector, can actually decay due to the annihilation of its constituents in this model.\footnote{The SM has an enhanced $U(1)_{L_e} \times U(1)_{L_\mu}$ lepton number symmetries which forbid the muonium to decay via annihilation. This is different from the dark sector we consider where the off-diagonal  nature of the dark photon coupling breaks the two $U(1)$ down to a diagonal $U(1)$ of dark lepton number. As a result, a dark muon is allowed to annihilate with a dark positron.} One way for the decay to occur is through its mixing with the dark photon, which could result in direct decays into a SM fermion pair. This decay rate, however, is suppressed by the small kinetic mixing parameter $\kappa$. (see footnote \ref{footnote1} for the upper bound on $\kappa$.) As a result, the other decay channel dominates, ${\rm Mu}_D\to 3 A'$, which only involves the dark gauge coupling but not $\kappa$.\footnote{We note that the ${\rm Mu}_D$ does not decay to two A' due to Furry's Theorem.} This is in close analogy to the $\Upsilon$ in the SM, which decays predominantly   into three gluons. 
As discussed above, here we assume a sufficiently large  $\kappa$ so that the dark photons will decay promptly into SM particles in the ATLAS or CMS detectors.\footnote{If the dark photon is long-lived, the leading process for its production is the mono-$A^\prime$ channel in the Higgs decay $h\to e_D^+ e_D^- A'$, as discussed earlier in section~\ref{sec:LHC}.}

\subsection{LHC Reach Using Lepton-jets}

In this subsection, we explore the multi-lepton final states from ${\rm Mu}_D$ production and decays as a signal of the dark sector at the LHC.

\begin{figure}[t]
	\centerline{\includegraphics[width=0.43\textwidth]{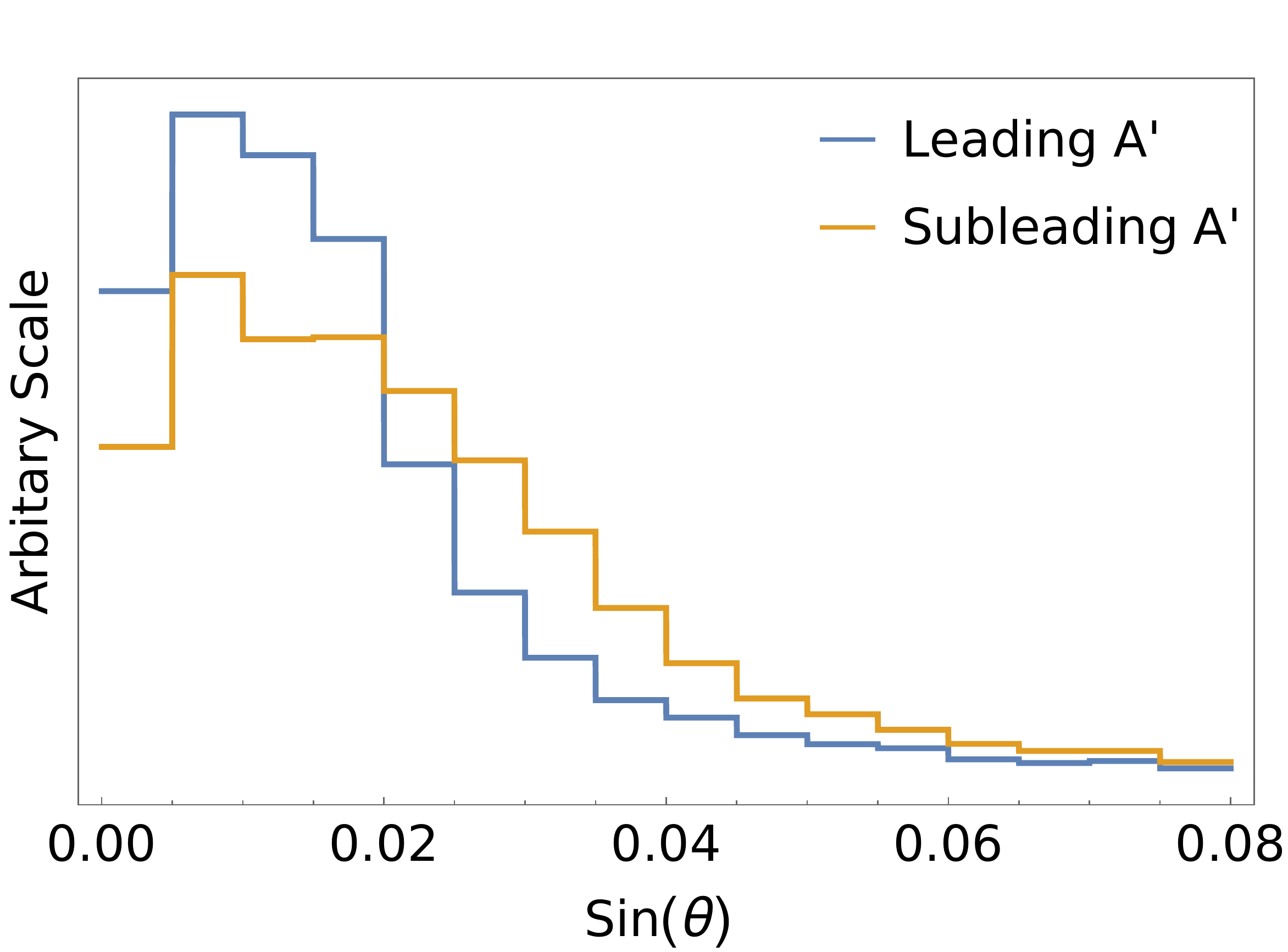}}
	\centerline{\includegraphics[width=0.43\textwidth]{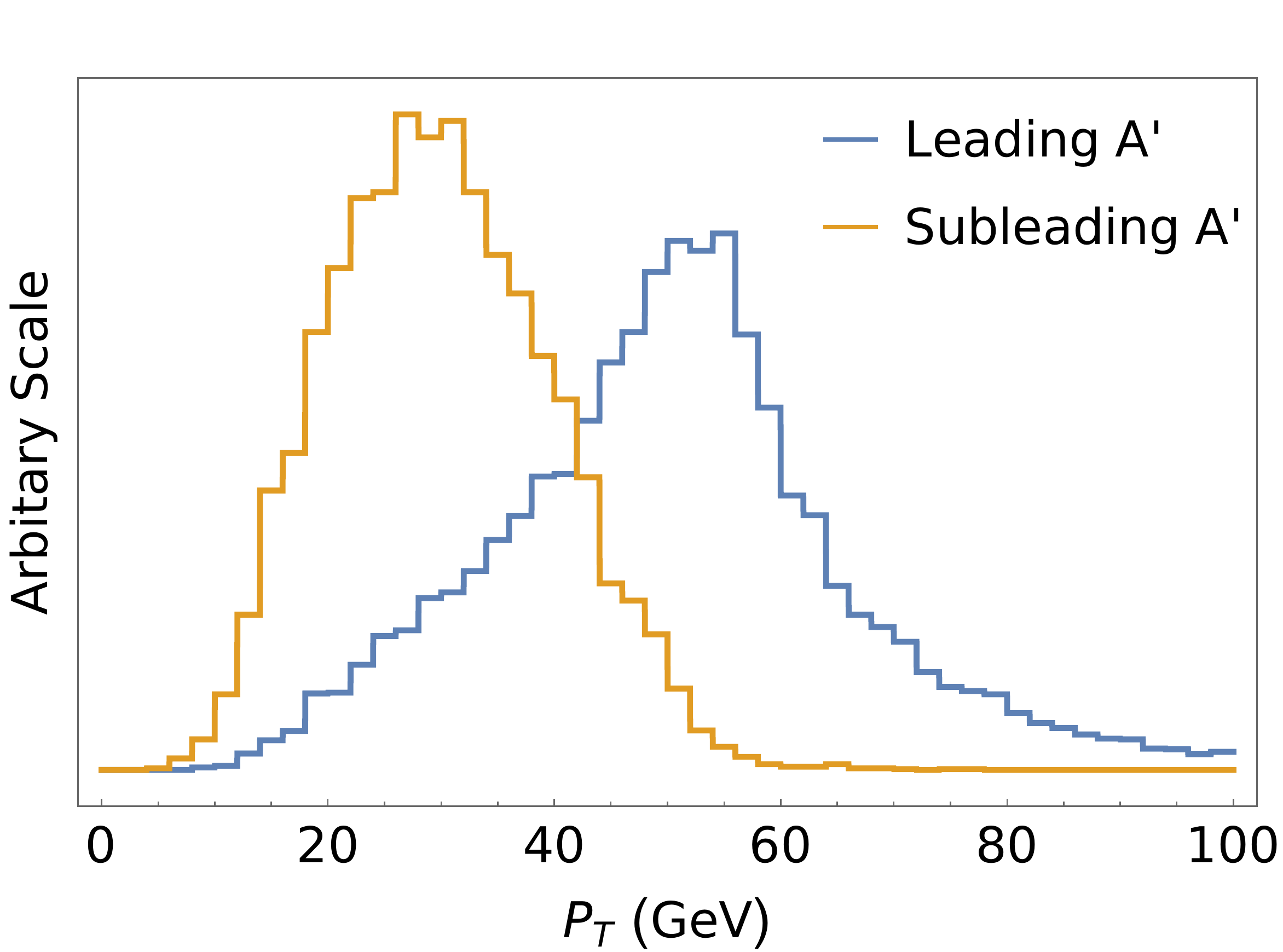}}
	\caption{Final state kinematic distributions of the process $h\to{\rm Mu}_D A' \to 4A'$ considered here, for dark photon mass $m_{A'}=0.5\,$GeV. The top plot shows the opening angle distribution of the charged lepton pairs from the dark photons with the highest (blue) and second-highest (yellow) $p_T$. The lower plots show the corresponding $p_T$ of the decaying dark photons.}\label{fig:openingangle}
\end{figure} 

We first create a {\tt FeynRules}~\cite{Alloul:2013bka} model file including the ${\rm Mu}_D$ and the dark photon, and use {\tt MadGraph}~\cite{Alwall:2011uj} to generate events for the Higgs boson production via gluon fusion and decay into ${\rm Mu}_D+A'$. The effective interaction between the Higgs boson, ${\rm Mu}_D$ and dark photon takes the contact interaction form, $c_{{\rm Mu}_D} h M_{\mu\nu} F'^{\mu\nu}$, where $M_{\mu\nu}$ and $F'^{\mu\nu}$ are the field strength tensors of ${\rm Mu}_D$ and dark photon, respectively, and

\begin{eqnarray}
c_{{\rm Mu}_D} = \frac{4 y_X g_{D}|\psi(0)|}{\sqrt{ m_{\chi}}(m^{2}_{h}+m^{2}_{A'}-4m^{2}_{\chi})} \ .
\end{eqnarray}
Each ${\rm Mu}_D$ from Higgs decay will decay dominantly into three dark photons. 
The analytic expression for this differential decay rate has been derived in~\cite{An:2015pva}. 
Here, for simplicity, we assume the decay occurs through the following operator $M_{\mu\nu}F'^{\mu\nu}F'_{\rho\sigma}F'^{\rho\sigma}$ and implement with {\tt FeynRules}. 
This is justified by studying the closely related decay in the SM, $J/\psi \to 3\gamma$, where it is found that the aforementioned operator is  a reasonable approximation over most of the decay phase space \cite{Adams:2008aa}.
Finally, we decay the dark photons into SM particles (lepton or quark pairs) via a kinetic mixing term with the photon. 
The resulting final states could contain as many as eight charged leptons at parton level.
The {\tt Madgraph} events are subsequently fed through hadronization using {\tt PYTHIA}~\cite{Sjostrand:2000wi} and detector simulation using {\tt DELPHES}~\cite{deFavereau:2013fsa}.

The eight-lepton final state has been proposed as the ``platinum'' channel  via a different mechanism in Ref.~\cite{Izaguirre:2018atq}, where limits were derived from searches for electroweakinos in supersymmetry in multilepton plus missing transverse energy (MET) final states \cite{Sirunyan:2017lae}. Our kinematics, as it turned out, is quite different from those studied in Ref.~\cite{Izaguirre:2018atq} and the electroweakino searches are not sensitive to our final states.

\begin{table}[t]
\begin{center}
	\begin{tabular}{ |c|c|c|c|c| } 
		\hline
		 & $8\ell$ & $6\ell$ & $4\ell$ & $2\ell$ \\ 
		 \hline		 
		a) & 5860  & 5177  & 4073 & 623 \\
		\hline 
		b) & 5466  & 4766  & 3655 & 554 \\ 
		\hline
		c) & 2056  & 998  & 367 & 11 \\
		\hline
		d) & 2056 & 998  & 367& 11 \\		
		\hline
		e) &1033  & 658 & 341 & 0  \\
		\hline
		f) &1013  & 656 & 341 &0  \\
		\hline
		Efficiency & 0.1013  & 0.0656  & 0.0341 &0  \\
		\hline
		\end{tabular}
	    \end{center}
\caption{Cut table showing number of events that pass different cuts. For each case, we generated 10,000 events to start with. These numbers correspond to model parameters $m_{\chi}$ = 20GeV and $m_{A'}$ = 0.5 GeV.
}
\label{table:1}
\end{table}

Instead we take advantage of the fact that the dark photon under our consideration is light, thus it is typically produced very boosted and the charged lepton pairs they decay into are collimated.
As a rough estimate, because four dark photons are produced in each decay, each dark photon carries energy $\sim m_h/4$. 
The opening angle between the charged lepton pair is then less than $4m_{A'}/m_h$. It is less than 0.1 for dark photon mass below a few GeV.
Indeed, this feature is shown in Fig.~\ref{fig:openingangle} with dark photon mass equal to 0.5\,GeV and dark matter mass equal to 20\,GeV.
We also show the corresponding leading and sub-leading dark photon $p_T$ distributions for the same simulated events which indicates that the charged leptons they decay into will be sufficiently energetic to pass the event selection.
Moreover, because only four dark photon are produced in each event, it is easy for the the charged lepton pair to be isolated with no other energetic object close by.
In this case, the lepton pair forms the spectacular object dubbed ``lepton-jet''~\cite{ArkaniHamed:2008qp, Baumgart:2009tn, Falkowski:2010cm} which is known to suffer from very low SM background.
Following Refs.~\cite{Sirunyan:2017lae, Baumgart:2009tn}, we impose the the following lepton-jet selection cuts:
\begin{enumerate}
\item[a)] For leading electron (muon) $p_{T} \geq 25\, (20)$GeV;
\item[b)] For sub-leading electron (muon) $p_{T} \geq 20\, (15)\,$GeV;
\item[c)] At least two additional charged leptons ($e$ or $\mu$) with $p_{T} \geq 10\,$GeV;
\item[d)] Missing transverse energy $\leq50\,$GeV.
\item[e)] We select events containing at least two lepton-jets. In each lepton-jet the opening angle less than 0.1;
\item[f)] In the annulus of $0.1 \leq R \leq 0.4$ around each lepton jet, the sum of hadronic $p_T$ should be less than 3 GeV.
\end{enumerate}

In table~\ref{table:1}, we show a cut flow table starting with signal events containing various numbers of charged leptons ($N_\ell = 2, 4, 6, 8$) at parton level, for a fixed set of model parameters, $m_\chi=$20\,GeV and $m_{A'}=0.5\,$GeV. Clearly, most of the signal events that pass the experimental cuts correspond to the $N_\ell=8$ final states.
The sum of numbers in the last row gives the overall efficiency factor $\epsilon$.

 \begin{figure}[t]
	\centerline{\includegraphics[width=0.48\textwidth]{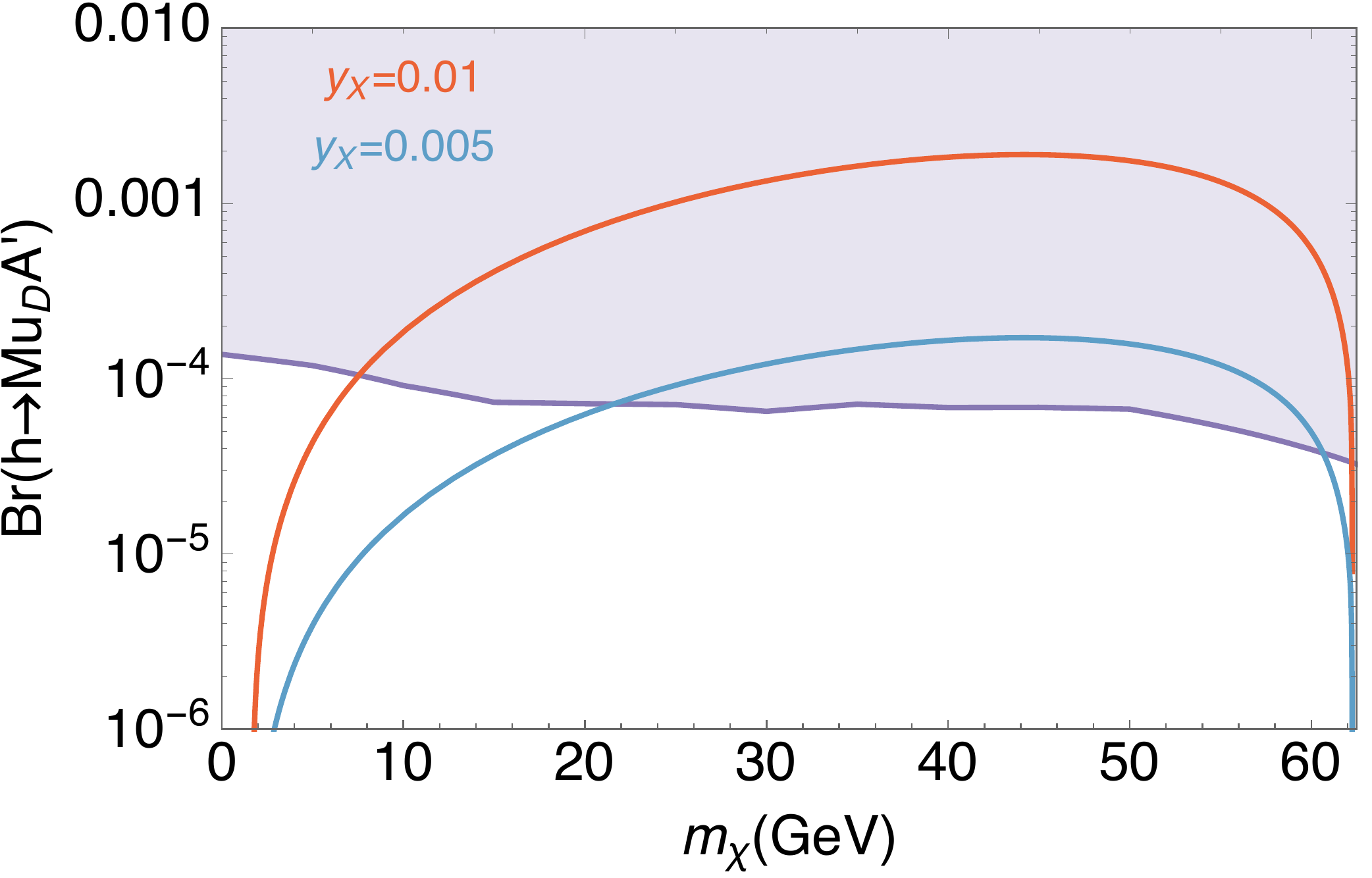}}
	\caption{Upper bound on the Higgs exotic decay branching ratio ${\rm Br} (h\to{\rm Mu}_D A')$ as a function of dark matter mass, derived using lepton-jet search described in the text with 36\,fb$^{-1}$ luminosity at the LHC. The other dark sector parameters are $\alpha_D=0.5$, $m_{A'}=0.5\,$GeV.}\label{fig:comp2}
\end{figure}

Fig.~\ref{fig:comp2} shows the 95\% C.L. limit on the upper bound of  the Higgs exotic decay branching ratio ${\rm Br} (h\to{\rm Mu}_D A')$, as a function of dark matter mass $m_\chi$, assuming 36\,fb$^{-1}$ luminosity at the LHC and a SM production rate for the Higgs boson. The excluded region is shown in the gray shaded region, which would produce more than 3 signal events, assuming negligible SM background. The sensitivity to this branching ratio gets slightly stronger for higher dark matter (thus higher dark muonium) mass. This feature could be understood because out of the four final state dark photons, three come from the dark muonium decay.
The heavier the dark muonium the more energetic these are photons are, and in turn it is easier for them to pass the lepton-jet selection cuts.
In this plot we fixed the dark photon mass to be 0.5 GeV. In our model the decay branching ratio into darkonium is determined by the dark Yukawa coupling and the dark matter mass, assuming a SM-like total width for the Higgs boson.\footnote{The decay width into the darkonium is too small to make an impact on the total width of the Higgs.} In the figure we also show in solid lines the decay branching ratio into the darkonium for two choices of dark Yukawa couplings, $y_X=0.01$ and 0.005, as a function of the dark matter mass. When the branching ratio is in the excluded region, the corresponding $y_X$ and $m_\chi$ is ruled out at the 95\% C.L.

\begin{figure*}
\centerline{\includegraphics[width=0.45\textwidth]{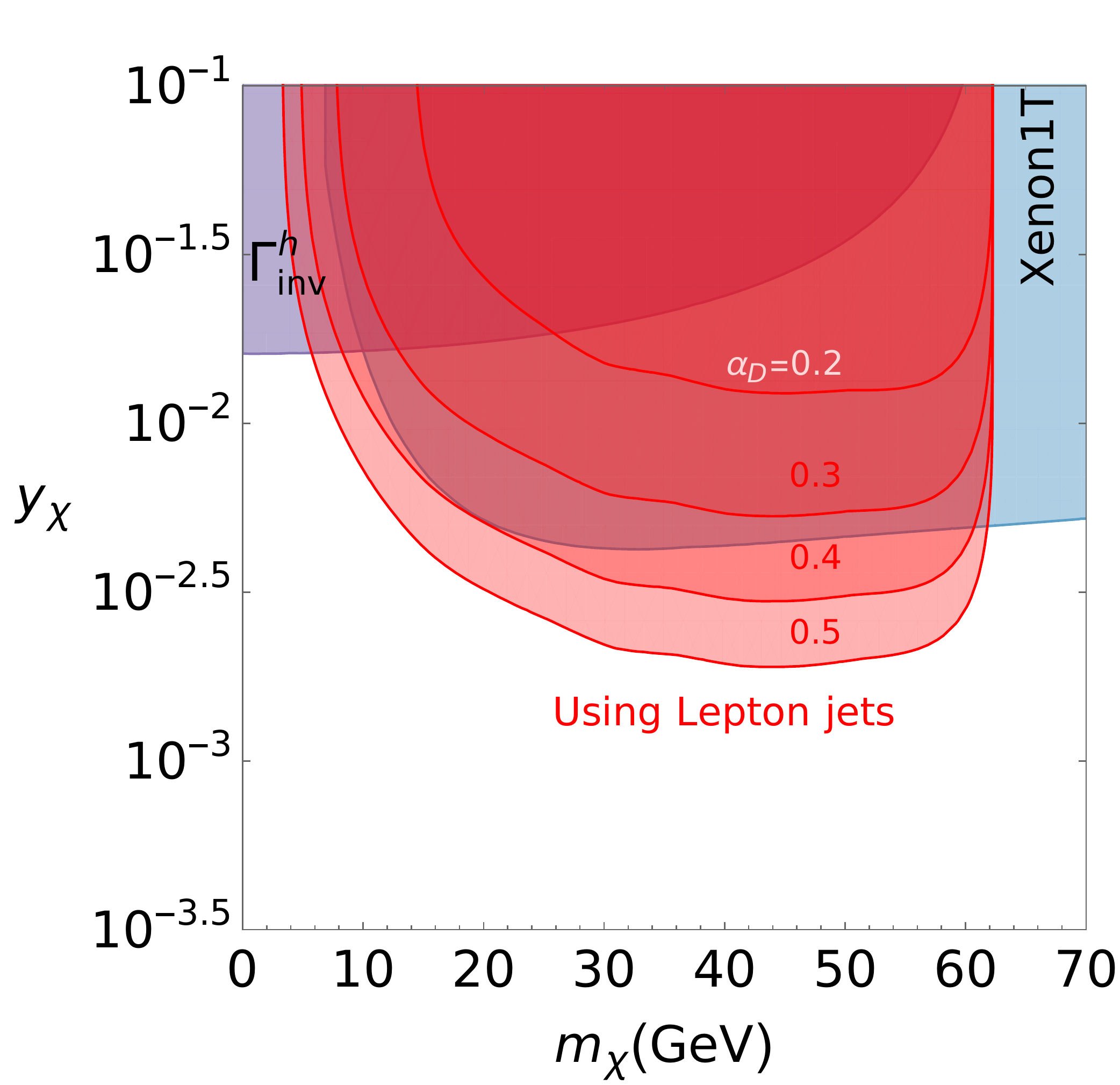} \hspace{0.5cm}
\includegraphics[width=0.45\textwidth]{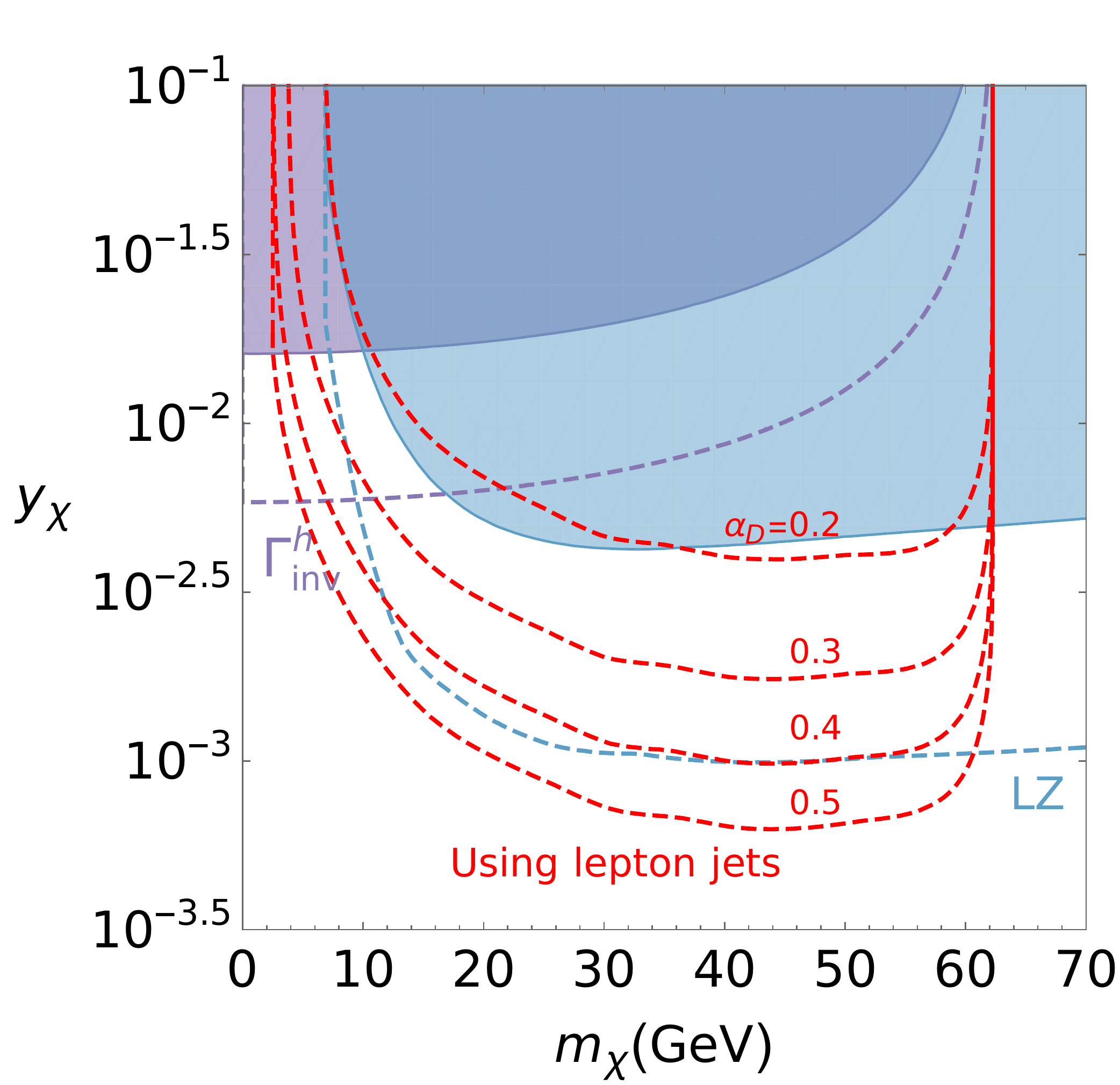}}
\caption{Lepton-jet search of Higgs exotic decay into the dark muonium channel as a probe of the Higgs portal to the dark sector. In the left plot, the red shaded regions correspond to 3 signal events (95\% CL reach in the absence of SM background) given the existing LHC luminosity, 36\,fb$^{-1}$. They are to be contrasted with the existing limits from dark matter direct detection and Higgs invisible decay searches. In the right plot, we show the future projection of various search approaches.
We hold the dark photon mass $m_{A'}=0.5\,$GeV and assume it decays promptly once produced at the LHC.
}\label{fig:Multi_Lepton}
\end{figure*}

In Fig.~\ref{fig:Multi_Lepton} (left), we present the above constraint in the $y_X$ versus $m_\chi$ parameter space (with $m_{A'}=0.5\,$GeV held fixed) for several values of $\alpha_D$, and compare it with the other existing limits from Higgs invisible decay searches and dark matter direct detection.
The lepton-jet analysis allows us to set a new useful limit on the model parameter space.
For $\alpha_D\gtrsim 0.3$, this limit beats the other constraints from direct detection and Higgs invisible decay.
In the right plot, we show the potential reach of the future high-luminosity runs of the LHC, with 3\,ab$^{-1}$.
Excitingly, it still remains competitive to the expected coverage of the future direct detection experiments such as LZ and XENONnT.
These two class of experiments  together will probe the territory with dark Yukawa coupling $y_X$ values as small as $\sim 10^{-3}$.

\begin{figure}[h]
	\centerline{\includegraphics[width=0.45\textwidth]{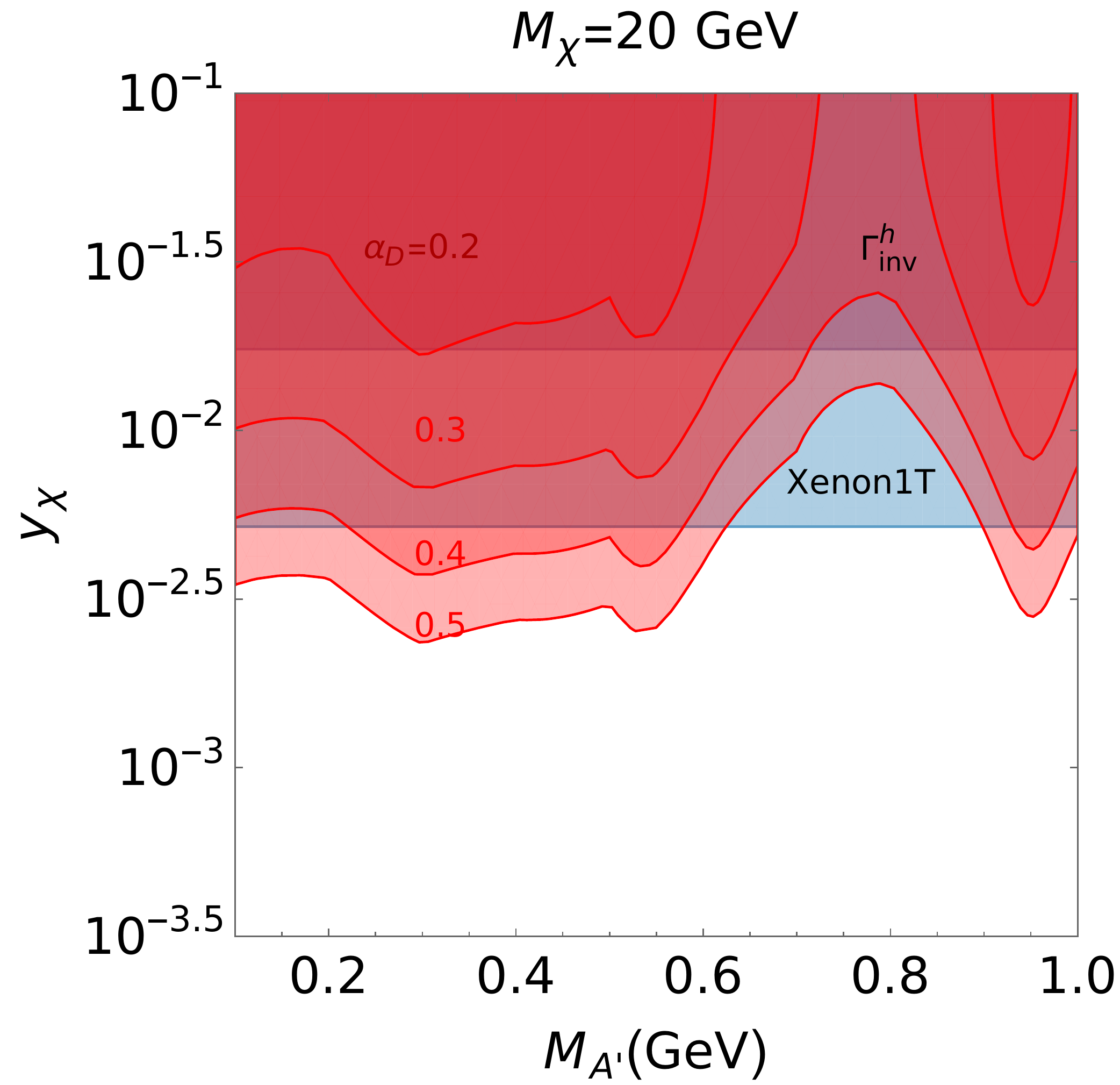}}
	\caption{Similar to Fig.~\ref{fig:Multi_Lepton} (left) but with constraints displayed in the $y_X$ versus $m_A'$ plane, with $m_\chi=20\,$GeV fixed. }\label{fig:yx-mAp}
\end{figure} 

In Fig.~\ref{fig:yx-mAp}, we fixed the dark matter mass (equal to 20 GeV) and plot the constraints in the $y_X$ versus $m_A'$ plane.
Here we only show the present limits. With large enough $\alpha_D$, we find that the lepton-jet search at LHC can set competitive limits for most of the parameter space with $m_{A'}$ less than a GeV, except for a window where the dark photon is nearly degenerate with the $\rho$ and $\omega$ mesons causing it decaying more often into hadrons (pions) instead of charged leptons. For dark photon heavier than GeV, this limit gets weakened quickly because 1) the opening angle of the charged lepton pair from dark photon decay becomes larger thus the efficiency for passing the lepton-jet cut lowers, and 2) the dark photon decay branching ratio into charged leptons decreases significantly because many hadronic decay channels open up.

\section{Conclusion}\label{sect:con}
In this work, we have introduced a dark photon to the minimal Higgs portal model which involves a fermionic dark matter coupled to a Higgs boson. 
The coupling between dark matter and dark photon allows for additional dark matter annihilation channels into the dark photons which helps to avoid the overclosure problem for dark matter masses below that of the Higgs boson.
At the same time, the dark photon exchange contributes to dark matter direct detection which in turn has important implications on its lifetime and the corresponding LHC phenomenology.
We explore two classes of models where the existing direct detection constraints are satisfied.
Depending upon the values of the kinetic mixing parameter $\kappa$ between the dark photon and the SM photon, we point out new opportunities for exploring both the energy and the lifetime frontiers. 

At the lifetime frontier, we study the mono-$A^\prime$ channel, where the Higgs boson decays to a dark photon plus MET. From the angular distribution of the emitted $A^\prime$, we realize that the recently proposed MATHUSLA experiment is a promising avenue to explore. Using its proposed detector setup, we are able to probe a complementary region of the parameter space as compared to dark matter direct detection experiments. While the direct detection experiments are adept at probing higher values of DM masses in the tens of GeV range, MATHUSLA is most useful for covering the lower DM mass range (a few GeV and below). 
This complementarity is shown by first plot in Fig.~\ref{fig:Long_Lived}.
Further, the MATHUSLA experiment could be sensitive to the dark photon mass up to tens of GeV, in cases where it is long lived due to a very small $\kappa$, as can be seen in the second plot in Fig.~\ref{fig:Long_Lived}. This region has not yet been covered by other terrestrial dark photon searches, supernova cooling, BBN, as well as dark matter direct detection experiments. MATHUSLA will be a unique probe of this region.

At the energy frontier, the process we consider involves a Higgs decaying to a dark muonium and a dark photon. The dark muonium further decays to three dark photons. 
In this scenario, since we are interested in low dark photon masses, the pair of charged leptons from its decay are highly collimated, leading to the striking lepton jet final states.
By exploring this final state, we find an interesting competition between the LHC and direct detection experiments such as Xenon1T and the future LZ experiment. Together, these experiments can probe Higgs-dark matter Yukawa coupling $y_{\chi}$ values as low as $10^{-3}$. 
Although lepton jets have been proposed as a signature of extended dark sectors before, a novel feature associated with the final states studied here is the absence of any significant MET. A dedicated experimental effort is therefore required. 

\acknowledgements

We acknowledge helpful discussions on the experimental aspects of lepton-jet searches with Walter Hopkins and Sasha Paramonov.   This work is supported in part by the U.S. Department of Energy under contracts No. DE-AC02-06CH11357 at Argonne and No. de-sc0010143 at Northwestern. This manuscript has been authored by Fermi Research Alliance, LLC under Contract No. DE-AC02-07CH11359 with the U.S. Department of Energy, Office of Science, Office of High Energy Physics. 

\bibliography{bibliography}
\end{document}